\documentclass{jfm}
\usepackage{xspace,colortbl}
\usepackage[square,comma]{natbib}
\usepackage{amssymb}
\usepackage{epsfig}
\usepackage{graphicx}
\usepackage{amsmath}
\usepackage{bm}
\input{tcilatex}

\begin{document}

\title[Crude sap ascent and tree recovery]
{The watering of tall trees  -\\   Embolization and  recovery}

\author[H. Gouin ]{Henri Gouin} \affiliation {Aix-Marseille Universit\'e, CNRS, Centrale Marseille, M2P2 UMR 7340  \\     13451  Marseille  France

  E-mails:\,\ henri.gouin@univ-amu.fr;\,\ henri.gouin@yahoo.fr
 \\

}

\maketitle
 \centerline{\textbf{Journal of Theoretical Biology,  369 (2015) 42–-50.}}

 \centerline{http://dx.doi.org/10.1016/j.jtbi.2015.01.009}

\noindent \textbf{Highlights }\\

 {$\bullet$ Biologists are still  debating tree recovery and the cohesion-tension theory.

$\bullet$ The concept of disjoining pressure is taken into account for high ascent of sap.

 {$\bullet$ Examples enable us to understand why the  embolized vessels can be refilled.

$\bullet$ The stability domain of liquid thin-films limits the maximum height of trees.  \\

\begin{abstract}

\noindent \textbf{Abstract:\ }
We can propound a thermo-mechanical understanding of the ascent of sap to the top of tall trees thanks to a comparison between experiments associated with the cohesion-tension theory and the disjoining pressure concept for liquid thin-films. When a segment of xylem is tight-filled with crude sap, the liquid pressure can be negative although the pressure in embolized vessels remains positive.   Examples are given that illustrate how embolized vessels can be refilled and why the ascent of sap is possible even in the tallest trees avoiding the problem due to cavitation. However, the  maximum height of trees is  limited by the stability domain of liquid thin-films.

\end{abstract}


\textbf{----------------------------------------------------------------------------------------------- }


\noindent \textbf{Introduction} \\

 The crude sap   ascends thanks to the negative pressure generated by the evaporation of water from the leaves.
  This  classical explanation of the sap ascent phenomenon in tall trees  is known as the cohesion-tension theory   (\cite{Dixon})   and is followed by a
quantitative analysis of the sap motion proposed by   \cite{vanderHonert}.
  The main experimental check  on the cohesion-tension theory comes from   the Scholander pressure chamber   (\cite{Scholander,Tyree}).\\
Trees   pose multiple challenges for crude sap transfer:\\
Conditions in the sap do not  approach the ultimate tensile strength of liquid water during transpiration (\cite{Herbert}).
   Nonetheless, the liquid water columns do break in  tracheary elements.
   Cavitation events in the xylem
   seem to have been acoustically detected with   ultrasonic transducers pressed
    against the external surface of the trees (\cite{Milburn},\cite{Tyran}). The porous  vessel walls can prevent the   gas bubbles from spreading and allow the flow to take alternate paths around the emptied segments (\cite{Mercury}).
The pores  connecting adjacent segments in the xylem vessels pass through the vessel walls, and are bifurcated by bordered-pit membranes which are thin physical fluid-transmitters. Pit membranes in pores are of fundamental importance  at nanometric scales;  applying the Laplace formula, the pressure difference across them can easily be of the order of 1 - 10 MPa (\cite{Meyra,Jansen}).   In the leaves and in the stems, the bordered-pit membranes serve as capillary seals that allow for a difference in pressure to exist between the liquid in the xylem and the gas phase outside (\cite{Tyree1,Sperry}).
The pressure in   water-containing neighbouring tracheids
may still be negative; a considerable pressure drop therefore exists
across the pit membranes (\cite{Choat}).
 No vessels are
continuous from roots to stems, from stems to shoots, and from shoots
to petioles, and  the water does not leave a
vessel in the axial direction  but laterally along a
 long stretch (\cite{0'Brien}).
\\
Consequently, trees seem to live in  unphysical conditions  (\cite{Holbrook1,Holbrook}),
 and  to be hydrated, they exploit
liquid water in   thermodynamically metastable states of negative pressure (\cite{Zwieniecki}).
At great elevation in  trees, the value of the negative pressure increases the risk of cavitation and
the formation of embolisms may cause a definitive break-down of the continuous column of sap, inducing  leaf death.
 For a negative pressure  $P_l =-0.6$ MPa in the
sap, corresponding to an approximate minimal value of the hydrostatic pressure for embolism reversal in plants, we obtain  a bubble radius $R \geq 0.24\, \mu \texttt{m}$  (\cite{Nardini});  then, when all the vessels are tight-filled, nucleation sites
naturally pre-existing in   crude water
 may spontaneously  embolize the tracheids (\cite{Pridgeon}).
Consequently, at high elevation, it does not seem  possible to refill a tube full of vapour
at a positive pressure when liquid-water must be at a negative pressure, but in the xylem, the
liquid-water's metastability - due to negative pressures -  may persist even in the absence of
transpiration. Once embolized vessels have reached a nearly full state,
is the refilling solution still at positive pressure, in mechanical equilibrium with some remaining air?\\

The most popular theory for the refilling process has been proposed by Holbrook  and Zwieniecki in several papers:  due to the fact that tracheary elements are generally in contact with numerous living cells (\cite{Zimm}), they hypothesized that crude sap is released into the vessel lumen from the adjacent living cells in a  manner similar to root exudation (\cite{Kramer}) and they assumed that the mechanism for water movement into embolized conduits involves the active secretion of solutes by the living cells. However, a survey across species indicated  the root pressure could reach 0.1-0.2 MPa above atmospheric pressure (\cite{Fisher}) and was the only logical source of embolized vessels' repairing  at night in smaller species with well-hydrated soil. The M\"{u}nch pumping mechanism (\cite{Munch}) was invoked, but basic challenges for this mechanism still persisted: osmotic pressures measured in sieve tubes do not scale with the height of a plant as one would expect (\cite{Turgeon,Johnson}) and such scenarios have not yet been empirically verified.
Hydraulic isolation was also required to permit the local creation of the positive pressures necessary to force the gas into solution yet the embolism removal might be concurrent with tree transpiration (\cite{Zwieniecki1}).
Additionally, refilling in the presence of tension in adjacent vessels required the induction of an energy-dissipating process that would locally pump liquid into the emptied vessels (\cite{Canny0}) or lower  the water potential in the vessel with the secretion of solutes (\cite{Zwieniecki}).
As a consequence,   \cite{Canny1,Canny,Johnson} and other authors (\cite{McCully}) suggested that alternative mechanisms might be required.
\\
Alternatively, for slightly compressible liquids, the molecular theory of capillarity, applied to liquid thin-films wetting solid substrates, demonstrates an unexpected behaviour
  in which liquids do not transmit the pressure to all their connected parts, as it is  for liquid-bulk parts   (\cite{Lifshitz}).
 Consequently, it  is possible to obtain an equilibrium between connected liquid parts where one  is at a positive pressure - the pressure in
 a liquid thin-film - and the other one  is at a negative pressure - the pressure in the liquid bulk.
 The vapour-gas phase in contact with the liquid thin-film is at the same positive pressure as the liquid thin-film. The refilling of xylem
   is not in contradiction with possible phase equilibria at different pressures in the stems.
 The   model associated with this behaviour corresponds  to the so-called \emph{disjoining pressure theory}.
\\

 The paper  analyses the results obtained in the  physical chemistry literature that are useful to explain  the refilling of tracheary elements and the watering  of tall trees.  The apparent  incompatibility between the model in  \cite{Gouin8}  and the cohesion-tension theory is now solved. The model allows to explain aspects of sap movement which the classical cohesion-tension theory was hitherto unable to satisfactorily account for, e.g. the refilling of the vessels in spring, in the morning  or after embolism events, as well as the compatibility with thermodynamics' principles.
\\
\emph{ The paper is organized as follows:}
 Section 1 is required by the fact that nanofluidic and liquid thin-films concepts are  fundamental physical tools for the rest of the paper. Following Derjaguin's Russian school of physical chemistry, we  propose an experimental overview of the disjoining pressure concept for liquid thin-films at equilibrium.
Thermodynamical potentials   are also recalled  for liquid thin-films.
 We end the section by a study of vertical motions  along   liquid thin-films: a comparison between
 liquid-motions' behaviours both in tight-filled  microtubes and in liquid thin-films  is proposed.
 It appears that slippage conditions on walls multiply the flow rate along liquid thin-films by an order ranging from $10^2$ to $10^4$ and
 consequently, liquid thin-films  flow-rate is not similar to Poiseuille's liquid-flow-rate. \newline
Section 2 is the most important - and completely new -  part of this research. The section focuses on trees containing  vessels considered as machines. From experiments presented in the previous section, a model of xylem    using liquid thin-films is proposed.
   Such a    model of xylem allows to explain both the thermodynamical  consistence of
   the cohesion-tension theory and the conditions of the crude-sap refilling     at high elevation.
This previous \emph{thought experiment}   is modified to take  account of air-vapour pockets:
 when the air-vapour pocket pressure is greater than the air-vapour bulk pressure, a huge flow occurs
 between the two parts filled by air-vapour gas to empty the air-vapour pockets although the liquid-bulk pressure is negative.\newline
Section 3 is a byproduct of  section 1. The section  shortly reproduces results we  previously published in the  literature and it is an important complement to Section 2. The   \emph{pancake-layer concept}, associated with the
breaking down of vertical liquid thin-films, allows to forecast the limit of validity of the model and yields a maximum   height for the tallest trees.   \newline
A conclusion ends the article;  this section suggests experiments  to  verify  the accuracy of sap  ascent   for tall trees and the accuracy of crude-sap's refilling.

\section{ The disjoining pressure for liquid thin-films}

\noindent The disjoining pressure is a physical concept specific to  liquid thin-films wetting  a flat solid surface and  bordered by a vapour  bulk.
 A complete description   is proposed by    \cite{Derjaguin}. Liquids in contact with solids are submitted to intermolecular
forces  making liquids a little compressible and  consequently heterogeneous; the stress tensor is non-spherical contrary to what it is in homogeneous bulks (\cite{gouin1}).

\subsection{ Horizontal liquid thin-films}
At a \emph{given temperature} $T_{_0}$, two experiments allow to understand the  physical meaning of  horizontal  liquid thin-films \emph{at equilibrium}:

 $\bullet$ The first experiment explaining the concept was carefully described in \cite{Derjaguin}:   a liquid
bulk   submitted to  pressure $p_{l_{b}}$
contains a microscopic bubble of radius $R$ contiguous to a solid
 (Fig. \ref{fig1}). The  bubble floats upward and approaches a horizontal smooth   plate, and a planar liquid thin-film is formed after some time.
\begin{figure}
\begin{center}
\includegraphics[width=8cm]{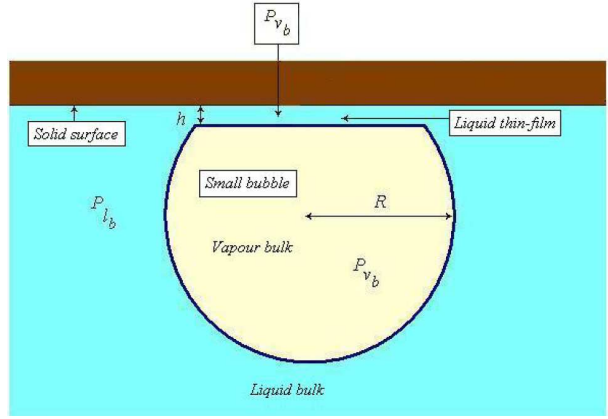}
\end{center}
\caption{\emph{The bubble method of determining the disjoining
pressure isotherms of wetting films.} The hydrostatic pressure in the liquid thin-film is the same as in the  microscopic bubble  and is different from the liquid-bulk pressure (from \protect\cite{Derjaguin},
page 330).} \label{fig1}
\end{figure}
The liquid thin-film separates the flat part of the bubble, which is squeezed onto
 the solid surface, from inside. Inside the bubble,  the pressure of the vapour bulk    is $p_{v_{b}}$.
The film is thin enough for gravity to be
neglected thickness-wise and the hydrostatic pressure of
the liquid thin-film  is  identical to the vapour-bulk pressure inside the bubble. Pressure $p_{v_{b}}$ differs from  pressure $p_{l_{b}}$ of the contiguous liquid
bulk. The  previous analysis can apply to the bulk pressure $p_{l_{b}}$ in the liquid a short distance away from the surface; the bulk pressure $p_{l_{b}}$ is not really affected by the gravity gradient because of the microscopic size of the bubble which remains spherical outside the liquid thin-film.  By using the   Laplace formula,   the difference between
the two bulk pressures is
\begin{equation}
p_{v_{b}}- p_{l_{b}} = \frac{2\gamma}{R},\label{Laplace}
\end{equation}
 where $\gamma$ is the surface tension   of the bubble liquid-vapour interface.
Pressure $p_{v_{b}}$ in   vapour bulk
$({v_{b}})$ of density $\rho_{v_{b}}$ (\emph{mother vapour-bulk}), and pressure $p_{l_{b}}$ in    liquid bulk   $(l_{b})$ of density
$\rho_{l_{b}}$ (\emph{mother liquid-bulk}), from which the liquid  thin-film extends, create  the   pressure difference already estimated in \eqref{Laplace}  and named $\Pi(h)$:
\begin{equation}
\Pi(h)  =p_{v_{b}}-p_{l_{b}}\, .  \label{disjoiningpressure}
\end{equation}
This    interlayer   pressure  $\Pi(h) $ additional to the mother liquid-bulk pressure is called
 the \emph{disjoining pressure} of the thin film of thickness $h$, and   curve\
 $h \longrightarrow\Pi(h)$\ \ - obtained by changing the bubble's radius   and thereby   film thickness $h$ -  is the
\emph{disjoining pressure isotherm}.
\\
   Derjaguin's clever idea was to create an analogy between liquid thin-films and liquid-vapour
interfaces of bubbles. Liquid thin-films - allowing  to obtain an equilibrium between
fluid phases   at different pressures - are physically similar to "bubbles' flat interfaces".
 The pressure in the liquid phase is different  from the liquid pressure in the liquid thin-film,
  which is the same as  the pressure in the vapour phase; thereby, the liquid
  - with a very small compressibility -  does not completely transmit  the  pressure in all places where it lays.

 $\bullet$   The second experiment illustrating the disjoining pressure concept is associated with
the  classical apparatus due to
 \cite{Sheludko} and is described in Fig.  \ref{fig2}.
\begin{figure}
\begin{center}
\includegraphics[width=9cm]{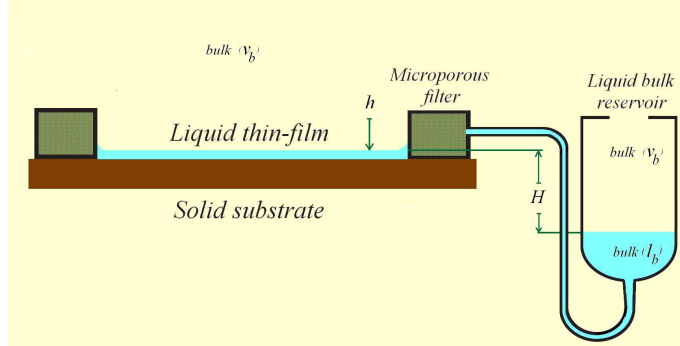}
\end{center}
\caption{\emph{Diagram of the technique for determining the disjoining
pressure isotherms of wetting thin-films on a solid substrate}. A circular wetting
thin-film is formed on a flat substrate to which a microporous filter with a
cylindrical hole is clamped. A pipe connects the filter filled with the
liquid to a reservoir containing the mother liquid-bulk that can be moved by
a micrometric device. The disjoining pressure is equal to $\Pi =(\protect%
\rho _{{l_{b}}}-\protect\rho _{{v_{b}}})\,gH$,  where $g$ is the gravity  acceleration  (from  \protect\cite{Derjaguin},
page 332).}
\label{fig2}
\end{figure}
   The
hydrostatic pressure in the liquid thin-film located between a solid wall and
a vapour bulk $(v_b)$ differs from the pressure in  contiguous liquid bulk $(l_{b})$.
\\
   As done in the first experiment,   the disjoining pressure is  equal to the difference
between     thin-film's interfacial surface pressure $p_{{v_{b}}}$ - which is the
pressure    of the mother  vapour-bulk  -  and mother liquid-bulk's
pressure $p_{{l_{b}}}$.  As explained below in Subsection 1.2,  the forces arising during the thinning of a film of uniform thickness
$h$ produce  disjoining pressure $\Pi (h)$ in the thin film according to the
surrounding bulks.

\subsection{ Chemical potential, Helmoltz's free energy and their consequences}
The chemical potential   of a fluid is a   versatile thermodynamical potential allowing
 to study  the equilibrium between vapour and liquid bulks. At a given temperature $T_{_0}$ - due to the pressure state-equation - the chemical potential of the fluid
  can be considered as a function of  fluid density $\rho$ (\cite{Rowlinson}). We choose - as reference chemical potential - $\mu_{_0}$ which is null in the  bulks
  of the phase equilibrium associated with the planar vapour-liquid interface:
  $$
  \mu_{_0}(\rho_v) =  \mu_{_0}(\rho_l) =0 ,
  $$
 where $\rho_v$ and $\rho_l$ are the vapour and the liquid bulk-densities, respectively.
The densities in the mother bulks are connected by the relation
 $$
  \mu_{_0}(\rho_{v_{b}}) =  \mu_{_0}(\rho_{l_{b}})
 $$
 and the pressure is
 $$
p(\rho) = \rho\, \mu_{_0}(\rho)- \Psi_{_0} (\rho) + p_{_0} ,
 $$
 where $p_{_0}$ is the pressure of planar liquid-vapour equilibrium, and due to thermodynamical relations (\cite{Callen}),
 $$
\Psi_{_0} (\rho) = \int_{\rho_v}^\rho \mu_{_0}(\tau)\, d\tau
 $$
 is the Helmoltz free energy per unit volume, where  in the integral, variable $\tau$    is the variable of integration. When the fluid is water, the
  Helmoltz free energy per unit volume  is classically called the water potential.
 \newline
 As described  in \cite{Derjaguin},\, Chapter 2\, and \cite{degennes2},\, Chapter 4,
  the thermodynamic potential  or Gibbs free-energy (per unit area)   of
the  liquid thin-film  is noted  $G$ and can be expressed as a function of
 $h$.
  In a reversible equilibrium change at constant temperature and constant pressures, accompanied by a change $dh$ in the interlayer thickness, the external forces do work  like $  - \Pi(h)\, dh$ and this work must be equal to the increment $dG$ of the thermodynamical potential. Hence,
  thanks to the relation
  \begin{equation*}
   \Pi(h) = - \left(\frac{\partial G}{\partial h}\right),
  \end{equation*}
  the partial derivative being taken at constant temperature and pressures, we can write $G$ in the form
\begin{equation*}
{G (h)} = \int_{h}^{+\infty} \Pi(\tau)\,d\tau,\label{Gibbs2}
\end{equation*}
where $h = 0$  is associated with a dry wall in contact with the mother vapour-bulk, $h = +\infty$ is associated with a wall in
 contact with  the mother liquid-bulk  when the value of  $G$  is equal to $0$;  variable $\tau$    is the variable of integration.   \\
 For liquid water   wetting the solid wall  constituted with different  substrates, experiments and calculations  are proposed
 in \cite{Sheludko}. The liquid thin-film thickness depends on the disjoining pressure value: for water at 20${{}^\circ}$Celsius
  and a disjoining pressure value of   an  order of several atmospheres, the  order of thickness is of some
  nanometres and   the liquid thin-film is a   nanolayer. In the liquid thin-film, the liquid is
inhomogeneous:  the variation of the liquid density is of the order of $4/100$ over a distance of some nanometres, but the density gradient
 is important (its order is  about  $10^{12}\ \texttt{kg.m}^{-4}$). \newline
    \begin{figure}
\begin{center}
\includegraphics[width=7cm]{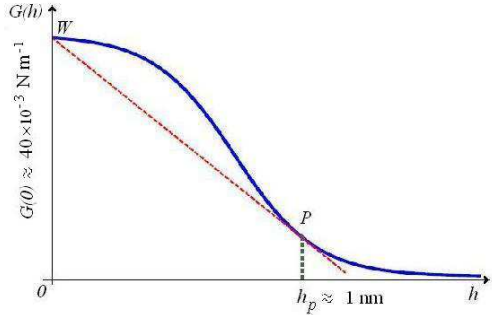}
\end{center}
\caption{{\emph{Sketch of the pancake layer thickness for water}. The construction of the tangent to  curve $G(h)$ issued from point
 \emph{W}  of coordinates  $(\emph{0},G(\emph{0}))$  yields point  \emph{P}; point
\emph{W } is associated with a high-energy surface  of the dry wall and point  \emph{P}  is associated with   pancake-thickness  $h_p$   (\cite{degennes2}}).   The values indicated for $h_p$ and $G(0)$ are approximatively those of water at 20${{}^\circ}$Celsius: the wetting film damping a solid wall is stable down to $h_p\approx 1 $ nm.}   \label{fig3}
\end{figure}
The coexistence of two adjacent liquid film-segments with different thicknesses is a phenomenon
  which can be represented by
 the equalities of   chemical potentials and   superficial tensions of the two films.
A spectacular case
  corresponds to the coexistence of a liquid thin-film of thickness $h_p$ and the dry solid wall.
    The liquid thin-film is the  so-called   pancake  layer  and corresponds  to the minimal thickness for
which a stable wetting film damps a solid wall.
 The minimal thickness  $h_p$ satisfies the relation
\begin{equation}
{G}(0) =  {G}(h_p) + h_p\, \Pi(h_ p). \label{pancake thickness}
\end{equation}
The geometrical
interpretation of Eq. (\ref{pancake thickness}) is proposed on Fig.  \ref{fig3}.

\subsection{ Vertical liquid thin-films}

The experiments in Subsection 1.1 and definitions in Subsection 1.2 can be   extended to vertical liquid thin-films
 wetting  a  vertical   solid surface (we propose a complete study in \cite{Gouin2}).
 We  denote  $x$ the  coordinate along the  upward-vertical direction   and $g$ the gravity acceleration.
 When altitude differences are of  one hundred meter order, the fluid densities in   mother bulks at level $x$, $({l_{b_x}})$ and  $({v_{b_x}})$
  are not perceptibly modified and  they are always denoted $\rho_{{l_{b}}}$ and $\rho_{{v_{b}}}$.  The vapour pressure is not perceptibly changed; the mother bulk $({v_{b_{_x}}})$ is the same as $({v_{b_{_0}}})$ and will be always denoted $({v_{b}})$.
  The  modifications are associated with a new chemical potential denoted $\mu_x$ at level $x$ and the liquid pressure,
 $$
 \mu_x =\, \mu_{_0} + g\,x,\qquad
 p_{l_{b_x}} \simeq\  p_{l_{b_{_0}}} - \rho_{{l_{b}}} g\, x\, .
 $$
   Consequently, the disjoining pressure strongly depends on the liquid thin-film altitude. The disjoining pressure can be considered as a function of $x$
   and if the disjoining pressure is zero at reference level, we obtain  (\cite{Gouin2}),
\begin{equation}
 \Pi(h_x)\simeq \,  \rho_{{l_{b}}} g\, x\,.\label{pressionaltitude}
\end{equation}
To make it clear, we draw a diagram on Fig.  \ref{fig4}. A reservoir along the
 solid wall is connected with the liquid thin-film at different levels.
  The liquid thin-film has a thickness $h_x$ depending on the altitude.
 At equilibrium the disjoining pressure   can interchangeably be  written
\begin{figure}
\begin{center}
\includegraphics[width=9cm]{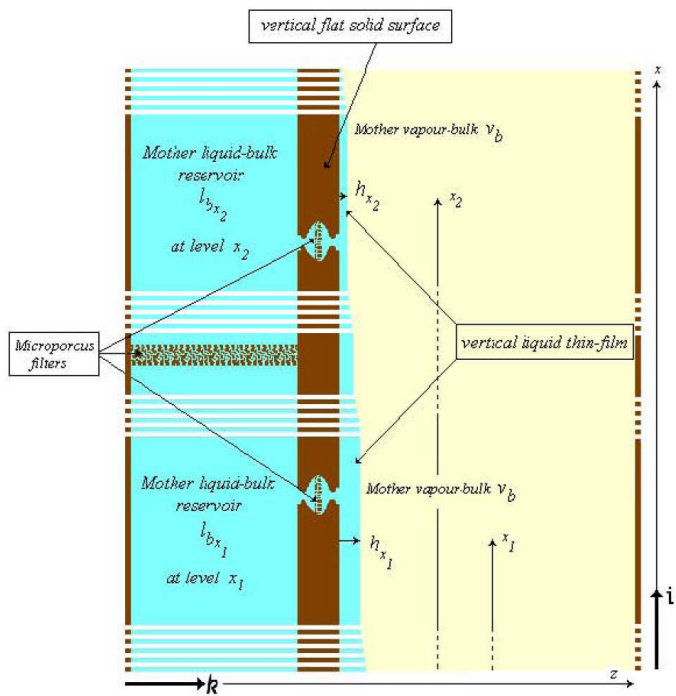}
\end{center}
\caption{\emph{Diagram of a vertical liquid thin-film}.  A   liquid thin-film bordered by a mother vapour-bulk $(v_b)$ wets
 a vertical flat solid surface. Mother liquid-bulk $(l_{b_x})$, but not $(v_{b})$,  depends on the x-level.  Like on Sheludko's apparatus,
microporous filters connect the mother-bulk reservoir at different altitudes with the liquid thin-film. The different
parts of the mother-bulk reservoir can    also be  connected by microporous filters.}
\label{fig4}
\end{figure}
\begin{equation}
 \Pi(h_x) =\, \Pi[x] \label{DPlevel}
 \end{equation}
 where  - for the sake of simplicity - $\Pi$ is a function of $x$ as indicated in
 \eqref{pressionaltitude}  as well as another function of $h_x$ as proposed in \eqref{disjoiningpressure}.
  Relation (\ref{DPlevel}) defines the connection between $x$-level and $h_x$-thickness.

\subsection{ Motions in the liquid thin-films}

The  results arising from molecular physics
allow to study the motions   of liquid thin-films of a thickness of some nanometers    (\cite{oron,Gouin7}).
There are qualitative experiments for slippage on  solid wall  when the film thickness
is of the mean free path  order  (\cite{Churaev}).
The boundary condition  on the wall  writes
\begin{equation*}
\boldsymbol{u}=L_{s}\, {d \boldsymbol{u}/{d n}},
\end{equation*}
where  $\boldsymbol{u}$ is the liquid velocity, ${d \boldsymbol{u}}/{d n}$
is the normal derivative at the wall and $L_{s}$ is the so-called \emph{Navier length}
 (\cite{Degennes}). The Navier length   may be as large as a few microns  (\cite{Tabeling}), and we obtained the  mean
liquid velocity  ${\mathbf{\overline{\boldsymbol{u}}}}$  along a vertical liquid thin-film  in  \cite{gouin5}\,:
\begin{equation}
  \nu\, {\mathbf{\overline{\boldsymbol{u}}}}=h_{x} \left( \frac{%
h_{x}}{3}+L_{s}\right) \left[ \,{\bf{grad}}\ \Pi (h_{x}) - g\,\boldsymbol{i}\,
\right] ,  \label{variation potentiel chimique}
\end{equation}
where $\nu$ denotes the kinematic viscosity and $\boldsymbol{i}$ the vertical unit vector
(we simply note that for   horizontal liquid thin-films,  $g = 0$). The increase in disjoining pressure comes from the decreasing thickness of the liquid thin-film creating the upward liquid motion.
The slippage condition multiplies the flow rate along liquid thin-films by a factor of $(1+3L_{s}/h_{x})$.
For example, if $h_{x}= 3\,\texttt{nm}$\,  and $%
L_{s}=100 \,\texttt{nm}$, which is a Navier length of small magnitude with respect to
experiments, the multiplicative factor is $10^2$; if $L_{s}= 7\, \mu \texttt{m}$, as
considered in \cite{Tabeling}, the multiplicative factor
 is $10^{4}$ which is of the same order as in nanotube calculations  (\cite{Gouin9,Garajeu}), and
  observations  (\cite{Mattia}).
\section{Embolization and recovery}

\noindent Since the beginning of the cohesion-tension theory, many efforts have   been done to understand sap motions in the vascular network and to replicate tree functions when vessels are under tension.  Synthetic systems simulating the transport processes have played an important role in  model testing, and methods creating microfluidic structures to mimic tree vasculature have been developed (\cite{Stone}); by using  synthetic hydrogel,  which plays the role of the pit membrane, \cite{Wheeler} captured some fundamental aspect of the xylem tension and flow.\\
Liquid  thin-films of some nanometre thickness can be considered as plane interlayers with respect to the diameters of capillary vessels, which  range from 10 to 500 $\mu$m. Adjacent xylem walls  are connected by  active bordered-pit membranes   with micropores (\cite{Meyra}). The  membranes   separate  two volumes of fluid, and generally refer   to  lipid bilayers that surround living cells or
intracellular compartments (\cite{Stroock}).  The micropores are a few tens of microns wide. Due to the meniscus curvatures   at  micropore apertures, marking off liquid-water bulk from air-vapour atmosphere, the water-bulk pressure is negative inside micropore reservoirs, but, surprisingly,  semi-permeable micropores allow flows of liquid-water at negative pressure  to be pushed toward air-vapour domains at positive pressure   (\cite{Tyree2}).\\
Water exits the leaves by evaporation through stomata into subsaturated air. Resistance of the stomata sits in the path of vapour diffusion between the interior surfaces of leaves and the atmosphere but many of the tallest trees appear to lack active loading mechanisms (\cite{Fu}). When active transpiration occurs, stomata are open and these pumps run. The growth and degrowth of bubbles is rapid within xylem segments, but at night, although the stomata are closed,   xylem vessels  developing embolies during the day   can be refilled with liquid-water and the metastability of the liquid-water may persist even in the absence of transpiration  (\cite{Zwieniecki1,Zufferey}).  It is interesting to note that optical measurements indicate capillary Young's contact-angles of about 50${%
{}^\circ}$ for water on the xylem at 20${%
{}^\circ}$Celsius {\footnote{This value is also an arithmetic average  of different Young angles proposed in the literature  (\cite{Mattia}).}}, suggesting that the xylem walls  are not fully wetting and the capillary spreading cannot really aid the liquid-water refilling but may explain the apparent segregation of liquid-water into droplets  (\cite{Zwieniecki2}).

\subsection{A diagram of  vessel elements for tall trees}

At usual temperature (for example 20${%
{}^\circ}$Celsius), we consider two vertical adjacent  vessels  linked
by micropore reservoirs at convenient negative pressures -  in the same physical conditions of Subsection 1.3 - and   with pit membranes   dotting their walls  (the bordered-pit membranes   of the order of several tens   of nanometres correspond to the microporous filters in Fig. \ref{fig4}).
The mother liquid-water bulk also  contains dissolved air and the mother
 vapor-bulk contains air  \footnote{An important property   of any  fluid mixture
  consisting of liquid water, its vapour
and  air allows to directly take     the results of Section 1 into  account (\cite{gouin4}).
The  mixture's total pressure is the sum
of  the partial pressures of the components, and at equilibrium  the
partial pressure of air is constant through
liquid-air and vapour-air domains. Consequently,   results from Section 1 are unchanged: the disjoining pressure of
the mixture is the same as for   fluid without  air
when a liquid thin-film only separates liquid and vapour bulks (\cite{Gouin2}).
}.\\
One  vessel - corresponding to subsaturated mother air-vapour bulk - is embolized with a  positive pressure;
 it generates a liquid thin-film   which wets the xylem wall (see  the upper part of Fig. \ref{fig5}).
The other vessel   is filled with the mother liquid-water bulk at a negative pressure linked to the liquid  thin-film
 thanks to a micropore reservoir with a bordered-pit membrane (see the lower part of Fig. \ref{fig5}).
 \begin{figure}
\begin{center}
\includegraphics[width=8cm]{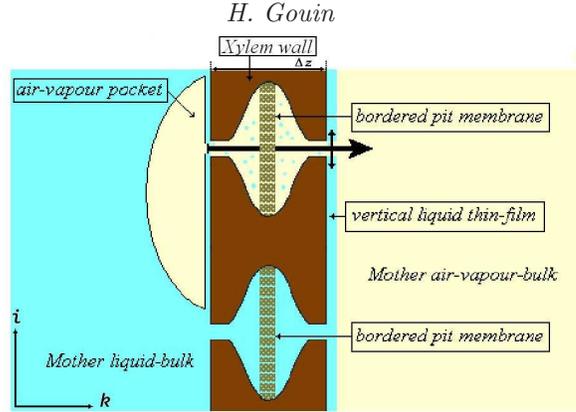}
\end{center}
\caption{At equilibrium between  bulks at altitude $x$, mother liquid-bulk $({l_{b_x}})$ balances the liquid thin-film and mother air-vapour-bulk $({v_{b}})$. It is not true for the vapour pocket and the mother air-vapour-bulk: when the pressure in the air-vapour pocket is greater than the pressure in mother air-vapour-bulk, the air-vapour pocket empties into the mother air-vapour-bulk.}
\label{fig5}
\end{figure}
Such a system can be in equilibrium, although the pressure is not the same in the two adjacent vessels.\\
In the same configuration, the  vessel elements are now assumed to be weakly out-of-equilibrium.
 As explained in Subsection 1.4, the driving force of
  the sap ascent comes from the decreasing thickness of the liquid thin-film wetting the walls in the embolized vessels;
   consequently the negative pressure value  of the mother liquid-water bulk in
    micropore reservoirs  decreases  (its absolute value increases). Additionally, air-vapour pockets can coexist with the mother liquid-water bulk of one of the
    two vessels. The air-vapour pockets  also generate  liquid   thin-films bordering  xylem walls (see Fig. \ref{fig5}).\\
  Due to their curvature,  the pressures of air-vapour pockets are generally higher than the mother  vapor-bulk pressure
in the other vessel. As proposed in the caption of Fig. \ref{fig5}, the vapour pockets and their
liquid thin-films   empty into the  vessel  with the lower air-vapour bulk pressure.
The analogy proposed in Subsection 1.1 between liquid thin-films and liquid-vapour interfaces of bubbles
allows to simply understand the  motion directions between air-vapour pockets and the mother air-vapour bulk: for example,
when two bubbles are included in a liquid bulk (corresponding to the mother liquid-water bulk at
 negative pressure),
the smallest bubble with the greatest pressure (i.e. the air-vapour pocket) empties into the largest bubble with the lowest pressure (i.e. the air-vapour bulk of the embolized vessel) (\cite{McCarthy}).

\noindent Such  events  happen  in particular at night, when - due to the absence of evaporation - the vapor is subsaturated in embolized vessels;
  the curvature of the air-vapor pockets generates a  pressure greater than the pressure in the embolized vessels.
  Conversely, during the day and strong sunlight, the vapour in  vessels is saturated
  by evaporation; the air-vapour pressure increases in the embolized vessels and the air-vapor gas must flow  back into the
  vessels with  air-vapour pockets of subsaturated vapour included in the mother liquid-water bulk at  negative pressure.

\subsection{Estimation of the transfer of  masses between  air-vapour pockets and  air-vapour  bulk }

It is interesting to estimate the magnitude of the air-vapour flow between air-vapour pockets and the mother air-vapour bulk.
At altitude $x$, the  mother bulks are still named $({l_{b_x}})$ and $({v_{b}})$, respectively and as indicated in footnote, the
disjoining pressure value has the same value as in Section 1.\\
At equilibrium between the two mother bulks, the vertical liquid thin-film - the liquid thin-film between  the mother air-vapour bulk and the xylem wall - is also in equilibrium
 and at the same pressure as  the mother air-vapour bulk. Depending on reservoir altitude $x$, the mother liquid-bulk can be at negative pressure. In that case, the pressure in the air-vapour pocket is \emph{assumed to be greater}
  than the  pressure in the mother air-vapour bulk. Consequently, the air-vapour pocket
   and the mother air-vapour bulk are not in equilibrium. Nonetheless, the  air-vapour pocket
   and the mother liquid-bulk may be at equilibrium, with a liquid thin-film separating
   the air-vapour pocket and the xylem wall; the liquid thin-film between
   the air-vapour pocket and the xylem wall is thinner than  the liquid thin-film between  the mother air-vapour bulk and the xylem wall.\\ In the case of perfect fluids,
   the equation of motion
   is deduced from \cite{serrin} and \cite{gouin6}.  Due to the fact that the chemical potential of the air-component
   is constant, we  get:
\begin{equation}
 \boldsymbol{a}+  {\rm grad}(\mu_{_0}+g\,x) = 0,  \label{motion1}
\end{equation}
where $\mu_{_0}$ is  the chemical potential of the water-component, not constant but depending on the density value and $ \boldsymbol{a}$ is the acceleration. As presented on Fig.  \ref{fig4}, distance $x$ corresponds to the ascending vertical height. Let us note that in the equilibrium case, Eq. (\ref{motion1}) writes
\begin{equation*}
 \mu_{_0}+g\,x = c_{_0}, \label{equilibrium1}
\end{equation*}
classically  corresponding to a connection between altitude $x$ and chemical potential  $\mu_{_0}$ ($c_{_0}$ is constant); in perfect motion, only acceleration term $ \boldsymbol{a}$ must be added to ${\rm grad}(\mu_{_0}+g\,x)$.
\\
Expansions of $\mu_{_0}$ at the first order in the  mother bulks  write  (\cite{Gouin2}),
 \begin{eqnarray}
  \mu_{_0} (\rho)& \approx & \frac{c_v^2}{\rho_v} (\rho -\rho_{v_b}), \qquad   \rm{in \   the\  air\texttt{-}vapour\  phase},\label{linearcp1}\\
    \mu_{_0} (\rho)& \approx & \frac{c_l^2}{\rho_l} (\rho -\rho_{l_b}), \qquad\   \rm{in \   the\   liquid\  phase},\label{linearcp2}
 \end{eqnarray}
where $c_v$ and $c_l$  are the isothermal sound velocity of the air-vapour mixture
and the liquid-water, respectively.
The  proofs of (\ref{linearcp1}) and (\ref{linearcp2}) are deduced from
$ {\partial \mu_{_0}}/{\partial\rho}\equiv (1/\rho)\left({\partial p}/{\partial\rho}\right),$
where $\partial p/\partial\rho$ are the values of the square of the isothermal sound velocity of the fluid at density $\rho$, taken in the vicinity of $\rho_{v_b}\approx \rho_{v}$ and $\rho_{l_b}\approx \rho_{l}$, respectively.
Consequently, the motion equations in the air-vapour mixture and in the liquid are:
 \begin{eqnarray*}
  \boldsymbol{a}& = & - \frac{c_v^2}{\rho_v} \,{\rm grad}\, \rho -
  g\, \boldsymbol{i},\qquad   \rm{in \   the\  air\texttt{-}vapour\  phase},\label{equacc1}\\
  \boldsymbol{a}& = & - \frac{c_l^2}{\rho_l} \,{\rm grad}\, \rho -
  g\,  \boldsymbol{i},   \qquad    \rm{in \   the\   liquid\  phase},\label{equacc2}
 \end{eqnarray*}
Components $\boldsymbol{a}. \boldsymbol{i}$ are equal in the two phases:   $\boldsymbol{a}. \boldsymbol{i}= -g \approx - 10\ \texttt{m}\, .\,  \texttt{s}^{-2}$.
Components $\boldsymbol{a}. \boldsymbol{k}$ are different:
 \begin{eqnarray}
   \boldsymbol{a}. \boldsymbol{k} & = &  - \frac{c_v^2}{\rho_v} \,\frac{\partial\rho}{\partial z} \approx - \frac{c_v^2}{\rho_v} \,\frac{\Delta\rho}{\Delta z},\qquad   \rm{in \   the\  air\texttt{-}vapour\  phase},\label{equakcc1}\\
 \boldsymbol{a}. \boldsymbol{k} & = & - \frac{c_l^2}{\rho_l} \,\frac{\partial\rho}{\partial z} \approx - \frac{c_l^2}{\rho_l} \,\frac{\Delta\rho}{\Delta z},   \qquad    \rm{in \   the\   liquid\  phase},\label{equakcc2}
 \end{eqnarray}
 where $\boldsymbol{k}$ is the normal-outward unit vector    to the vertical wall and $z$ is the associated coordinate.
  In \eqref{equakcc1},  $\Delta\rho$ represents the difference of densities  between the air-vapour pocket  and the mother air-vapour  bulk
  and in  \eqref{equakcc2}, $\Delta\rho$ represents  the difference of densities between  the liquid thin-film squeezed between the air-vapour pocket and the vertical liquid thin-film;  $\Delta z$
     is an estimation of the distance between the air-vapour pocket and the
air-vapour  bulk.\\
The physical values of the isothermal sound velocity, $c_v \approx 3 \times 10^2 \
 \texttt{m}\  \texttt{s}{^{-1}}$ and $c_l \approx 1.5 \times 10^3 \ \texttt{m}\ \texttt{s}^{-1}$,
 are given in the \cite{Handbook}.   We assume the following conditions to be corresponding to a pressure in the air-vapour pocket slightly greater than the mother air-vapour pressure:
$$\Delta z \approx 10^{-3}\  \texttt{m},
\quad \Delta \rho /\rho_v   \approx 10^{-2},\quad \Delta \rho /\rho_l \approx 10^{-3}.$$
We get:
\begin{eqnarray*}
  \boldsymbol{a}. \boldsymbol{k}& \approx & 10^{6}\, \texttt{m}\   \texttt{s}^{-2}, \qquad\qquad   \rm{in \   the\  air\texttt{-}vapour\  phase},\label{equakcc3}\\
\boldsymbol{a}. \boldsymbol{k} & \approx & 2  \times10^{7}\, \texttt{m}\   \texttt{s}^{-2}, \ \qquad    \rm{in \   the\   liquid\  phase}.\label{equakcc4}
 \end{eqnarray*}
We obtain  an outsized acceleration
for air-vapour gas - between the air-vapour pocket and   its associated liquid thin-film -
 directed towards the mother air-vapour bulk.
 The air-vapour pocket empties into the mother air-vapour bulk.
 One notes that the acceleration in the liquid thin-film is  an order higher
 than the acceleration in air-vapour gas.  The liquid thin-film associated with
  the air-vapour pocket    empties faster than the air-vapour pocket itself.\\
  Conversely, when the  mother air-vapour pressure is slightly greater than the   air-vapour pocket pressure, the air-vapour bulk can embolize the liquid tight-filled  vessel elements with opposite acceleration.
  \\
  It is interesting to notice,  when  $\Delta z \approx 10^{-3}\  \texttt{m}$\, and\, $ \boldsymbol{a}. \boldsymbol{k}  \approx 10^{6}\, \texttt{m}\  \texttt{s}^{-2}$,  we obtain a transfer time  $t \approx  4.4\times10^{-5}\texttt{s}$ corresponding to the ultrasonic frequency $\omega =t^{-1} \approx 20\, \texttt{kHz}$. The fast accelerations of air-vapour gas through micropores can thus generate ultrasounds and may explain
the acoustical measurements obtained in experiments (\cite{Milburn,Tyran}).\\
 The viscosity and diffusion prevailing inside  pit membranes
and into micro-channels substantially decrease   the acceleration value previously estimated
in the case of perfect motions.  The magnitude of the viscosity of simple wetting
fluids increases  when they are
confined between solid walls,  and there is a direct correlation between the air-seeding threshold and the pit pore membranes' diameters (\cite{Jansen}).
However, the acceleration magnitude is so large for perfect motions that it remains very important for
viscous fluids and   semi-permeable bordered-pit membranes.

\subsection{ Some remarks on motions in tight-filled vessels and in liquid-water thin-films}

It is interesting to compare  the behaviour of liquid motions both in tight-filled vessels
and   liquid-water thin-films.\\
 When the vessel elements are  tight-filled with crude sap, the liquid-water motions
  are  Poiseuille flows  (\cite{Zimm}).
 The Poiseuille flow is \emph{rigid} and the pressure effects are   propagated
onto the vessel walls.\newline
When the vessel elements are embolized, liquid-water thin-films damp the xylem walls.
Equation (\ref{variation potentiel chimique}) governs the liquid motion along
the thin-film,  and as we have seen in Subsection 1.4, allows to obtain a non-negligible flow rate.
The flow rate can increase or decrease  depending on the local
disjoining pressure  value.  The tree's  versatility allows it to adapt to the disjoining
pressure gradient effects by opening or closing the stomata and the curvature of pit pores, so that the bulk pressure in
micropores can be more or less negative and so, the transport of water in tight-filled vessels is differently
dispatched through the stem parts. \\
It is noteworthy that embolized vessels fundamentally contribute to the crude-sap ascent and to the refilling of the tree. Thus,   as indicated by   \cite{Zimm}, it is not surprising  that the heartwood may contain liquid under positive pressure while in the sapwood the transpiration stream moves along a gradient of negative pressure.  Embolized vessels creating liquid-water thin-films with non-negligible  flows may provide an important contribution to  tree refilling   by watering the heartwood. The  heartwood liquid-water  may also fill the vessel elements at negative pressure through the bordered-pit membranes connecting the thin-films and the mother liquid-bulk at negative pressure.\\
It is noticeable that if we replace the flat surfaces of the vessels (at molecular scale)   with corrugated surface,  it is much easier to obtain the
complete wetting requirement, which is otherwise only partial, as indicated at the beginning of the  section;  thus,  trees can avoid having very high
energy surfaces. However, they are still   internally wet  if crude sap flows  through
wedge-shaped corrugated pores. The wedge does not have to be perfect on the
nanometric scale to significantly enhance the amount of liquid flowing at modest pressures, the walls    being
considered as  plane surfaces endowed with an average surface energy.\\
To be efficient for  sap transportation, the tubes' diameters should be as wide as possible; because of the micron  size of the xylem tubes, this is not the case.
Consequently, the  tracheary elements' network must be important. As the sap movement is  induced by  the transpiration across
micropores located in tree leaves and the transpiration is bounded by  micropores' sizes, it seems natural to surmise
that the diameters of vessels must  not be too large to generate a sufficient sap movement. \\
Recent advances in tree hydraulics have demonstrated that, contrary to what was previously believed, embolism and repair may be far from routine in trees.
Trees can recover partially or totally from the deleterious effects of water stress until they reach a lethal threshold of cavitation (\cite{Delzon}).  This result can be related with the fact that thin films with a thickness greater than the pancake-layer's one are highly stable; this behaviour is different from bubble stability, which is associated with a saddle point (\cite{Slemrod}).

\section{ Topmost trees}
\noindent The main reason of the
  maximum size that trees can reach is  not well understood (\cite{Koch}).
 Amazingly,  the above studies for thin films allow to estimate a boundary value for thin-films' altitude. The boundary value corresponds to the limit of validity of the disjoining pressure concept.  Other mechanical or biological constraints may suggest adaptation to height-induced costs, but our model nevertheless limits the maximum height of trees.
The tallest trees are not the ones with the largest demand for tension; it is  rather dry climate shrubs that demand it. This  observation seems to be in accordance with  the possible existence of thin-films in embolized vessels at high elevation.   For tall trees, we have seen  that   liquid thin-films and disjoining pressure
may indirectly contribute to the xylem refilling.
The thickness of the liquid thin-film decreases when its altitude  increases.
The xylem refilling is not possible anymore  when the liquid thin-film breaks down.
  The liquid thin-film disrupts when thickness  reaches the pancake layer thickness-value.
  The pancake layer of liquid thin-films was presented in Subsection 1.2;
our aim is  to point out a numerical simulation connected with  physical data of xylem  such that previous results provide a
value of maximum height for a vertical water thin-film wetting  xylem wall. \\
\noindent  We consider water at $20 {%
{}^\circ}$Celsius. The
main physical coefficients are
 \begin{figure}
\begin{center}
\includegraphics[width=5.8cm]{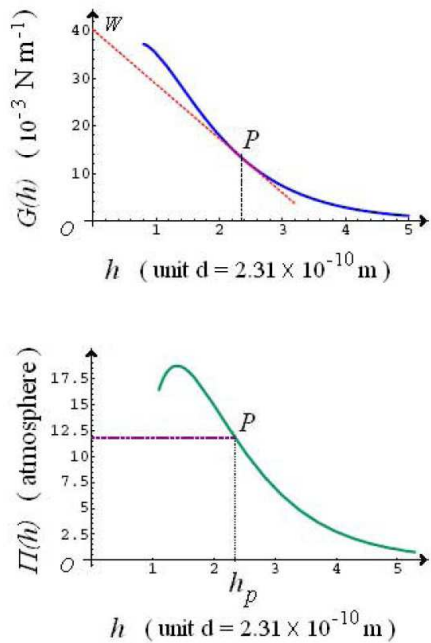}
\end{center}
\caption{\footnotesize \emph{{The maximum height of trees.}}   \emph{Upper graph}:  $G(h)$-graph.  The unit of $x$-axis graduated by $h$  is $%
d=2.31\times 10^{-10}$ \texttt{m} ; the unit of $y-$axis is $10^{-3}$
   N m$^{-1}$.   {\emph{Lower graph}: $\Pi(h)$-graph. The unit of $x$-axis graduated by $h$
is $d=2.31\times 10^{-10}$ \texttt{m}\,; the unit of $y-$axis is one
atmosphere  (From \cite{gouin5}, page 21).
}}
\label{fig6}
\end{figure}
$\rho_l =998$ kg\ m$^{-3}$, $c_l = 1.478\times 10^{3}$ m\ s$^{-1}$, $\gamma =  72.5 \times 10^{-3}$ N m$^{-1}$ (\cite{Handbook}).
As  indicated in Section 2, the Young contact angle  between the xylem wall and the liquid-vapour water interface is $\theta \approx 50{{}^\circ}$.   Other necessary physical coefficients due to molecular interactions are proposed in \cite{Gouin2,Gouin9} and used for the numerical calculations and
we can compare calculations obtained for trees with the sketch of the pancake layer thickness presented in Fig. \,\ref{fig3}:\\
In the upper graph of Fig.\ \ref{fig6} we present the free energy graph  $G(h)$ associated with trees' xylem walls.
Due to $h>({1}/{2})\,\sigma_l$ - where \ $\sigma_l=2.8\, \times 10^{-10} $ m\  \ stands for the water molecular diameter (\cite{Israel}) -   point \emph{W} is obtained by an
interpolation associated with the concave part of the  \emph{G}-curve. Point  \emph{P}\  follows from the drawing of the tangent line issued from
  \emph{W}   to the $G$-curve. In the lower graph of Fig.\,\ref{fig6},
  we present the disjoining pressure graph $\Pi(h)$. The real physical part of the disjoining pressure graph corresponds to $
\partial\Pi/\partial h <0$. The hypothetic part,  where $%
\partial\Pi/\partial h >0$, is also  obtained by  \cite{Derjaguin}.
The reference length
is of the same order as $\sigma_l$. The total pancake thickness $h_p$ is about
one nanometer order corresponding to a good thickness value for a
high-energy surface  (\cite{deGennes}); consequently in the tall trees, at high level, the thickness of the liquid thin-film must be of a few
nanometers. Point \emph{P} on  the lower graph corresponds to point \emph{P} on upper graph. The lower graph is theoretically obtained in \cite{gouin5},
and is in accordance with experimental curves obtained in the literature. Let us note that the crude sap is not pure water; its vapour-liquid surface tension
has a lower value than the surface tension of pure water   and it is possible to obtain the same spreading coefficients
with less energetic surfaces. \newline
When $x_{_{P}}$ corresponds to the altitude of the pancake layer,  $\Pi(h_p) \simeq \, \rho_l\,g\,x_{_{P}}$.  At this
altitude, we must approximatively add 20 meters corresponding to the ascent of sap due to capillarity and osmotic pressure, and reading on the lower graph of Fig.\ \ref{fig6}, we obtain  a maximum film height
of approximatively $140$ meters ($20+120$ meters) corresponding to $12$ atmospheres, which is of the same level order as the topmost trees, as a giant, 128 meter-tall
 eucalyptus or a 135 meter-tall sequoia (\cite{Flindt}).

\section{ Conclusion}

\noindent
Theoretical developments in the field relating to  the disjoining pressure  could  explain sap rise relatively to the cohesion-tension theory and also explain how sap might rise notwithstanding the existence of discontinuities or embolisms in the tracheary elements.   The disjoining pressure of liquid thin-films being the exhaust valve allowing for crude sap ascent, the embolized vessels constitute a necessary network for the watering and recovery of  tall trees  (\cite{Lampinen} argued that embolisms were necessary for the ascent of sap).
The molecular forces   can create liquid-water thin-films on the walls of xylem when the micropore pressures are versatilely adapted thanks to pit membranes. Nevertheless, simple in vivo observations  of the nanometre thickness of the liquid-water thin-films   are not easy to implement  and the  direct measurement difficulties  prevent their detection. The progression of MEMS technology (\cite{Tabeling}), and tomography  (\cite{Herman}), may provide a new route towards this goal.\\
As done in \cite{Stroock}, we ask questions:  "if plants can do it, why don't we? Why do human technologies not use liquids under tension?"\ We can conclude that if
these biophysical considerations are experimentally verified,  they would seem to prove that trees can be an example to use technologies for liquids under tension connected with liquids in contact with solid substrates  at nanoscale range.
They would provide a context in which nanofluid mechanics
points to a rich array of  biological physics and future technical challenges.
\\

{\footnotesize \textbf{Acknowledgements:} I am grateful to  the editor and the three anonymous reviewers for suggestions and comments during the review process.}

\newpage

 \noindent\textbf{Extended introduction  and additive comments removed from the Journal of Theoretical Biology.}\\

 The aim of the paper is to analyze the results obtained in the theoretical physical chemistry literature that are useful to explain  the watering  of tall trees in  the ascent of crude sap and the refilling of xylem microtubes. All the calculations already obtained are not performed again but only directly presented together with  the  bibliography explaining the detailed calculations   (\cite{Derjaguin,gouin1,Gouin7,Gouin8,Gouin2,gouin5,Gouin9}), both for the sake of simplicity and to make the demonstrations   more accessible. In this paper, we have to note that the apparent  incompatibility between the model in  \cite{gouin1}  and the cohesion-tension theory is solved. The model allows to explain aspects of sap movement which the classical cohesion-tension theory was hitherto unable to satisfactorily account for, e.g. the refilling of the vessels in spring, in the morning, or after embolism events as well as the compatibility with thermodynamics' principles.\\

\emph{A general view of the watering of  trees.}\\

Trees are engines running on  water, and on carbon  dioxide when exposed to sunlight. These circumstances pose multiple challenges for crude sap transfer  (\cite{Holbrook,Stroock}):
unlike animals, plants miss an active pump to move liquids along their vascular system. The  crude sap - mainly liquid water  absorbed and transported through trees - ascents thanks to the negative pressure generated by the evaporation of water from the leaves.
Additively, trees operate a second vascular system - phloem sieve tubes - for the circulation of metabolites though their living tissues,  with positive pressures, and elaborated sap flows passing from leaves to roots.
Measurements of the pressure within the terminal xylem vessels illustrate an
 extraordinary consequence of trees' behaviour for moving liquid water up to
  their leaves:  the pressure must be negative and the liquid water is under
  tension although  it seems prone to cavitation.\\ Trees do not  approach the ultimate tensile strength of liquid water during transpiration (\cite{Herbert}).
   Nonetheless, the liquid water columns do break in  xylem microtubes.
   Multiple types of measurements provide evidence for the cavitation of
   the liquid water in the xylem:   cavitation events in the xylem microtubes
   seem to have been acoustically detected with   ultrasonic transducers pressed
    against the external surface of the trees (\cite{Milburn,Tyree}). The porous  vessel walls can prevent the   gas bubbles from spreading and allow the flow to take alternate paths around the emptied segments (\cite{Mercury}).
The principal flow of water during transpiration goes to evaporation through stomata on the underside of the leaves.  The pores - or bordered pits - connecting adjacent segments in the xylem vessels pass through the vessel walls, and are bifurcated by bordered-pit membranes which are thin physical fluid-transmitters. Pit membranes in pores are of fundamental importance  at nanometric scales;  applying the Laplace formula, the pressure difference across them can easily be of the order of 1 to 10 MPa (\cite{Meyra}). They play a crucial role in maintaining the integrity of the water transport system  (\cite{Jansen}).  When wetted on both sides, the bordered-pit membranes allow the liquid-water flow to pass through. These membranes seem to allow for unusual  transport processes in trees:  flows of liquid water at negative pressures in the xylem counter flows of  elaborated sap at positive pressure in the phloem. In the leaves, these membranes serve as capillary seals that allow for a difference in pressure to exist between the liquid in the xylem and the gas phase outside. In the stem, the bordered-pit membranes also serve as seals between a gas-filled segment and an adjacent  liquid-filled segment, and avoid the propagation of massive embolisms when isolating the air-filled xylem tubes from the rest of  the xylematic stem (\cite{Tyree1}).
Consequently, trees seem to live in  unphysical conditions (\cite{Zwieniecki}),
 and  to be hydrated, they exploit
liquid water in   thermodynamically metastable states of negative pressure.\\
Xylem microtubes constitute the crude-sap watering network.
Crude sap    contains diluted salts
  but  its physical properties are roughly those of liquid water.
 In any tree, the crude sap is driven along xylem microtubes; consequently,
  hydrodynamics,  capillarity,  and osmotic pressure
  yield a crude sap ascent  of  a  few tens of meters  only (\cite{Zimm}).\\
As explained before, the pressure in   the water-storing tracheids of leaves can be strongly
 negative and consequently the pressure in the xylem microtubes of stems can also remain strongly negative  (\cite{Canny}).
This  classical explanation of the sap ascent phenomenon in tall trees  is known as the well-known cohesion-tension theory  propounded in
1893-1895 by \cite{Boehm},  \cite{Dixon} and \cite{Askenasy}, followed by a
quantitative analysis of the sap motion proposed by   \cite{vanderHonert}.
  According  to this theory, the crude sap  tightly  fills   microtubes of dead xylem cells and its transport is due to a gradient of negative pressure  producing the traction necessary to lift water against gravity. A main experimental checking  on the cohesion-tension theory comes from an apparatus called   Scholander pressure chamber   (\cite{Scholander}):  a leaf attached to a stem is placed inside a sealed chamber; compressed air is  slowly added to the chamber. As the pressure increases to a convenient level, the sap is forced out of the xylem and is visible at the cut end of the stem. The   required pressure is  opposite and of equal magnitude to the water pressure in the water-storing  tracheids in the leaf.
The decrease in the negative pressure is related to the closing of the aperture of microscopic stomata in leaves through which water vapour is lost by transpiration.
\\
\cite{Haberlandt}  described water-storing tracheids in leaves; they
are roundish in shape, and located either at the tips of the veins or
detached from transporting xylem. In more recent papers they
have been called   tracheid idioblats   (\cite{Foster}). The
spacing considered in  \cite{Pridgeon}, is about $2\, \mu \texttt{m}$ or less at the top of trees as suggested on Fig.\,2 of the paper by   \cite{Koch}, which is the right  size to
prevent  cavitation for nucleus germs of the same
order of magnitude.\\
  No vessels are
continuous from roots to stems, from stems to shoots, and from shoots
to petioles. The vessels do not all run neatly parallel and form a
network  generally  up to a few centimeters long. The
ends usually taper out; it is very important for the understanding
of water conduction to realize that the water does not leave a
vessel in axial direction through the very end but laterally along a
relatively long stretch where the two vessels, the ending and the
continuing  ones, run side by side.
The vascular bundles of some leaves are surrounded
by a bundle sheath, containing a suberized layer comparable to the one
of the Casperian strip in the roots  (\cite{0'Brien}). This seal
separates the apoplast into two compartments, one inside and the
other one outside the bundle sheath. The two areas are only connected by the
plasmodesmata that connect living cells.
The pressure in the intact, water-containing neighbouring tracheids,
may still be negative; a considerable pressure drop therefore exists
across the pit membranes and the large variations in their structure enhance their important role in xylem transport (\cite{Choat}). Pressure chamber measurements cannot be
considered as pressure values of the stem xylem without special
precautions, simply because they are taken elsewhere.
Hydraulically then, the leaf is very sharply separated from the
stem. The wet wood area of elms appears to act as  a single, giant
osmotic cell that is separated   from
the sapwood area by a semi-permeable  membrane. This can be visualized somewhat like a Traube
membrane, as early plant physiologists called it  (\cite{Traube}).\\

\emph{Problems associated with the cohesion-tension theory and the water recovery  in xylem microtubes.}\\

 Nonetheless, several objections question and  seem  to challenge the validity of the cohesion-tension theory, and worse, to preclude the possibility of refilling embolized xylem tubes.\\
We first refer  to
the well-known book by \cite{Zimm}. He said:
 "The heartwood is referred to as a wet wood. It may
contain liquid under positive pressure while in the sapwood the
transpiration stream moves along a gradient of negative pressures.
Why is the water of the central wet core not drawn into the sapwood?
Free water, \emph{i.e.} water in tracheids, decreases in successively older
layers of wood as the number of embolized tracheids increases. The
heartwood is relatively dry \emph{i.e.} most tracheids are embolized.
 It is rather ironic that a wound
in the wet wood area, which bleeds liquid for a long period of time,
thus appears to have the transpiration stream as a source of water,
in spite of the fact that the pressure of the transpiration stream
is negative most of the time.
It should be quite clear by now that a drop in xylem pressure below
a critical level causes cavitations and normally puts the xylem out
of function permanently. The cause of such a pressure drop can be
either a failing of water to the xylem by the roots, or excessive
demand by transpiration."\\
Following these comments, at great elevation in  trees, the value of the negative pressure increases the risk of cavitation and
 consequently, the formation of embolisms may cause a definitive break-down of the continuous column of sap inducing  leaf death.
In \cite{Gouin8} we  take another consideration into account: crude sap is a fluid with a superficial tension $\gamma$ lower
than the superficial tension of pure water, which is about $72.5 \times 10^{-3}$ N m$^{-1}$ at
20${{}^\circ}$Celsius as indicated in \cite{Meyra}; if we consider a microscopic air-vapour
bubble  with a diameter $2 R$ smaller than
xylem microtube diameters, the difference between   air-vapour pressure $P_v$
and  liquid sap pressure $P_l$ is expressed by the Laplace formula
$\displaystyle P_v-P_l = 2\,\gamma/R$ (\cite{Bruhat}); the air-vapour pressure is
positive and
 consequently unstable
bubbles will appear  when $\displaystyle
  R\geq
- 2\,\gamma/P_l$. For a negative pressure  $P_l =-0.6$ MPa in the
sap, corresponding to an approximative minimal value of the hydrostatic pressure for embolism reversal in plants of \emph{Laurus nobilis} (\cite{Nardini}), we obtain  $R \geq 0.24\, \mu \texttt{m}$;  then, when all the vessels are tight-filled, germs
naturally pre-existing in   crude water
 may spontaneously  embolize the tracheids.\\
 Another   objection
to the  confidence  in the
cohesion-tension theory was also the  experiment by
 \cite{Preston}  who demonstrated that tall trees survived double saw-cuts, made through the cross-sectional area of
the trunk to sever all xylem elements, by
overlapping them. This result,  confirmed   by
several authors (e.g.
 \cite{Mackay,Sperry}), does not seem to be in agreement with
the possibility of strong negative pressures in the water-tight microtubes.
Using a xylem pressure probe, \cite{Balling} showed that, in many
circumstances, this apparatus does not measure any water tension
 (\cite{Tyree}). However, there are other possibilities  for the tree's survival (researchers presented some physical evidences for the local refilling that restores embolized conduits by visualizing the conduits with microscope (\cite{Canny0,Canny1,McCully,Holbrook1}).\\
A negative argument seems  also to come from the crude sap's recovery in embolized xylem tubes.
At high elevation in a tree, it does not seem  possible to refill a tube full of vapour-gas
at a positive pressure when liquid-water must be at a negative pressure. In the xylem, the
liquid-water's metastability - due to negative pressures -  may persist even in the absence of
transpiration. Consequently, refilling processes pose a  tough physical challenge to push the
liquid-water back into the xylem vessels: once embolized vessels have reached a nearly full state,
is the refilling solution still at positive pressure, in mechanical equilibrium with some remaining air?\\

\emph{Theories of xylem embolism repair.}\\

The refilling process poses a physical challenge and theoretical problems.\\
The most popular theory has been proposed by Holbrook  \& Zwieniecki in several papers.  Due to the fact that xylem microtubes are generally in contact with numerous living cells (\cite{Zimm}), they hypothesize that crude sap is released into the vessel lumen from the adjacent living cells in a  manner similar to root exudation (\cite{Kramer}) and they assume that the mechanism for water movement into embolized conduits involves the active secretion of solutes by the living cells. Nonetheless, a survey across species indicated that the root pressure can reach 0.1-0.2 MPa above atmospheric pressure (\cite{Fisher}) and was the only logical source of embolized vessels' repairing  at night in smaller species with well-hydrated soil. The M\"{u}nch pumping mechanism is invoked but basic challenges for the M$\rm\ddot{u}$nch mechanism still persist (\cite{Munch}): osmotic pressures measured in sieve tubes do not scale with the height of a plant as one would expect (\cite{Turgeon}) and such scenarios have not yet been empirically verified.
Hydraulic isolation is also required to permit the local creation of the positive pressures necessary to force the gas into solution and the embolism removal may be concurrent with tree transpiration (\cite{Zwieniecki1}).
Additively, refilling in the presence of tension in adjacent vessels requires the induction of an energy-dissipating process that locally pumps liquid into the emptied vessels (\cite{Canny0}) or lowers the water potential in the vessel with the secretion of solutes (\cite{Zwieniecki}.
As a consequence,   \cite{Canny0} and other authors suggested that alternative mechanisms were required.

Alternatively, for slightly compressible liquids, the molecular theory of capillarity, applied to liquid thin-films wetting solid substrates, points out an unexpected behaviour
  in which liquids do not transmit the pressure to all their connected parts, as it is well known for liquid-bulk parts   (\cite{Lifshitz}).
 Consequently, it  is possible to obtain an equilibrium between connected liquid parts where one  is at a positive pressure - the pressure in
 a liquid thin-film - and the other   is at a negative pressure - the pressure in the liquid bulk.
 The vapour-gas phase in contact with the liquid thin-film is at the same positive pressure as the liquid thin-film.
  \newline The experiments and model associated with this behaviour correspond  to the so-called \emph{disjoining pressure theory}  (\cite{Derjaguin,deGennes}).
  The disjoining pressure theory allows for the equilibrium between liquid-water-bulk parts
  with negative pressures and vapour-gas-bulk parts with positive pressures. The cohesion-tension theory thus appears
  to be compatible with the laws of thermodynamics and   molecular physics, and the refilling of xylem tubes by the crude sap
   is not in contradiction with possible phase equilibria at different pressures in the stems.  \\

\end{document}